\documentclass{sig-alternate}
\usepackage{graphicx}
\usepackage{amsmath}
\usepackage{amsfonts}
\usepackage{amssymb}
\usepackage{latexsym}

\newcommand{\YA}{Yahoo Answers }
\newcommand{\YAc}{Yahoo Answers}
\newcommand{\comment}[1]{}

\newtheorem{hypothesis}{Hypothesis}
\newtheorem{lemma}{Lemma}

\begin{document}

\title{Harvesting Collective Intelligence}
\subtitle{Temporal Behavior in Yahoo Answers}

\numberofauthors{3} 
%
\author{
%
%
\alignauthor
       Christina Aperjis\\
       \affaddr{Social Computing Lab}\\
       \affaddr{HP Labs}\\
       \email{christina.aperjis@hp.com}
\alignauthor
        Bernardo A. Huberman\\
       \affaddr{Social Computing Lab}\\
       \affaddr{HP Labs}\\
       \email{bernardo.huberman@hp.com}
\alignauthor
       Fang Wu\\
       \affaddr{Social Computing Lab}\\
       \affaddr{HP Labs}\\
       \email{fang.wu@hp.com}
}

\maketitle

\begin{abstract}

When harvesting collective intelligence, a user wishes to maximize the accuracy and value of the acquired information without spending too much time collecting it.  We empirically study how people behave when facing these conflicting objectives using data from \YAc, a community driven question-and-answer site.
We take two complementary approaches.

We first study how users behave when trying to maximize the amount of the acquired information, while minimizing the waiting time.
We identify and quantify how question authors at \YA trade off the number of answers they receive and the cost of waiting.  We find that users are willing to wait more to obtain an additional answer when they have only received a small number of answers; this implies decreasing marginal returns in the amount of collected information.  We also estimate the user's utility function from the data.

Our second approach focuses on how users assess the qualities of the individual answers without explicitly considering the cost of waiting.  We assume that users make a sequence of decisions, deciding to wait for an additional answer as long as the quality of the current answer exceeds some threshold.  Under this model, the probability distribution for the number of answers that a question gets is an inverse Gaussian, which is a Zipf-like distribution.  We use the data to validate this conclusion.

\end{abstract}

\section{Introduction}

When searching for an answer to a question, people generally prefer to get both high quality and speedy answers.  The fact that it usually takes longer to find a better answer creates a speed-accuracy tradeoff which is inherent in information seeking.  Stopping the search for information early gives speed, while stopping the search for information late gives accuracy.  This paper studies how people behave with respect to this tradeoff.  A natural setting to study the speed-accuracy tradeoff is \YAc.

\YA is a community-driven question-and-answer site that allows users to both submit questions to be answered by the community and answer questions posed by other users.
With more than 21 million unique users in the United States and 90 million worldwide, it is the leading social Q\&A site.

At \YAc, users post questions seeking to harvest the collective intelligence of others in the system.  Once a user submits a question, the question is posted on the site.  Other users can then submit answers to the question, which are also posted on the site.  When a question author is satisfied with the answers he received, he closes the question, and thus terminates his search for answers.  The user then uses information from the answers he received to build his own final answer to his question.

There are two aspects that users value with respect to the final answers they obtain: {\em accuracy} and {\em speed}, and thus they try to maximize the accuracy of their final answers without waiting too long. The accuracy of the final answer depends on the accuracy of all individual answers that the question received.

Anyone posting a question faces the following tradeoff at any given point in time.  He can either build his final answer or wait.  If he waits, he may achieve a higher accuracy in the future, but also incurs a cost for waiting.  The user wishes to build his final answer at the optimal stopping time.
We take two complementary approaches to study users' behavior with respect to this stopping problem.

Our first approach studies the speed-accuracy tradeoff by using the number of answers as a proxy for accuracy.  In particular, we assume that the user approximates the accuracy of his final answer by the number of answers that his question gets.  Thus, the user faces the following tradeoff: he prefers more to less answers, but does not want to wait too long.  We analyze \YA data to identify and quantify this tradeoff.  Our first finding is that users are willing to wait more to obtain one additional answer when they have only received a small number of answers; this implies  decreasing marginal returns in the number of answers.  Formally, this implies a concave utility function in the amount of information.  We then estimate the utility function from the data.

Our second approach considers the qualities of the individual answers without explicitly computing the cost of waiting.  We assume that users 
decide to wait as long as the value of the current answer exceeds some threshold.  Under this model, the probability distribution for the number of answers that a question gets is an inverse Gaussian, which is a Zipf-like distribution.  We use the data to validate this conclusion.

The rest of the paper is organized as follows.  In Section \ref{sec:related} we review related work.  In Section \ref{sec:rules} we describe \YA focussing on the rules that are important for our analysis.  In Section \ref{sec:tradeoff} we empirically study the speed-accuracy tradeoff by using the number of answers as a proxy for accuracy.  In Section \ref{sec:quality} we focus on how users assess quality.  We conclude in Section \ref{sec:conclusion}.

\section{Related Work}
\label{sec:related}

This paper studies behavior with respect to stopping when people face speed-accuracy tradeoffs using \YA data.  Three streams of research are thus related to our work: (1) behavior with respect to stopping, (2) speed-accuracy tradeoffs, and (3) empirical studies on \YAc.  These are briefly discussed below.

The behavior of people with respect to stopping problems has been studied extensively in the context of the secretary problem~\cite{classical}.
In the classical secretary problem applicants are interviewed sequentially in a random order, and the goal is to maximize the probability of choosing the best applicant.  The applicants can be ranked from best to worst with no ties.  After each interview, the applicant is either accepted or rejected.
If the decision maker knows the total number of applicants $n$, for large $n$ the optimal policy is to interview and reject the first $n/e$ applicants (where $e$ is the base of the natural logarithm) and then to accept the next who is better than these interviewed candidates~\cite{classical}.

Experimental studies of both the classical secretary problem and variants show that people tend to stop too early and give insufficient consideration to the yet-to-be-seen applicants (e.g., ~\cite{bearden}).  On the other hand, when there are search costs and recall (backward solicitation) of previously inspected alternatives is allowed, people tend to search longer than the optimum~\cite{recall}.
We note a key difference with the setting of information seeking: while in the secretary problem only one secretary can be hired, an information seeker can combine information from multiple sources to build a more accurate answer for his question.  Moreover, in the secretary problem the decision maker does not face a speed-accuracy tradeoff because time does not affect his payoff.

The speed-accuracy tradeoff has been considered in various settings.  One example is a setting where a group cooperates to solve a problem~\cite{speed_accuracy}.
In psychology, on the other hand, the speed-accuracy tradeoff is used to describe the tradeoff between how fast a task can be performed and how many mistakes are made in performing the task (e.g., \cite{psycho}).

There has been a number of empirical studies that use data from \YA and other question answering communities.
Data from \YA have been used to predict whether a particular answer
will be chosen as the best answer~\cite{adamic}, and whether a user will be satisfied with the answers to his question~\cite{datasetA}.
Content analysis has been used to study the criteria with which users select the best answers to their questions~\cite{selection_criteria}.
Shah et al. study the effect of user participation on the success of a social Q\&A site~\cite{QAparticipation}.
Aji and Agichtein analyze the factors that influence how the \YA community responds to a question~\cite{datasetB}.
Finally, various characteristics of user behavior in terms of asking and answering questions have been considered in~\cite{koutrika}.
To the best of our knowledge, there have been no studies that consider user behavior in terms of the speed-accuracy tradeoff in either \YA or any other question answering communities.

\section{Yahoo Answers}
\label{sec:rules}

In this section we describe \YA focussing on the rules that are important for our analysis.  We also briefly describe the data we use.

\YA is a question-and-answer site that allows users to both submit questions to be answered and answer questions asked by other users.
Once a user submits a question, the question is posted on the \YA site.  Other users can then see the question and submit answers, which are also posted on the site.  According to standard \YA terminology, the user that asks the question is called the {\em asker}, and a user answering is called an {\em answerer}.  In this paper we study the behavior of the asker, and thus the word user is used to describe the asker.

Once the user starts receiving answers to his question, he can choose the best answer at any point in time.  After the best answer to a question is selected, the question does not receive any additional answers.  We thus say that a user closes the question when he chooses the best answer.  Closing the question is equivalent to terminating the search for answers to the question.

We expect that the user is satisfied with the answers he received, when he closes the question.  The user then uses information from these answers to build his own final answer to his question.
Throughout the paper, we use the term {\em final answer} to refer to the conclusion that the question author draws by reading the answers to his question.  The final answer is {\em not} posted on the \YA site, and is often not recorded.

Questions have a 4-day open period.  If a question does not receive any answers within the 4-day open period, it expires and is deleted. However, before the question expires, the asker has the option to extend the time period a question is open by four more days.  The time can only be extended once.
In both datasets, most questions are open for less than four days (96 hours).  Throughout the paper we only consider questions that were open for less than 100 hours.

If the asker does not choose a best answer to his question within the 4-day open period, then the question is up for voting, that is other \YA users can vote to determine the best answer to the question.

For the purposes of this paper we only consider questions for which the best answer was selected by the asker.  The reason is that we are interested in the time that the asker terminates his search for information by closing the question.  If the asker selects the best answer, this is the time that the best answer was selected.  On the other hand, if the asker does not select a best answer, we have no relevant information (we do not know when and whether the asker built his final answer).

In this paper we use two \YA datasets.  Dataset A consists of 81,832 questions and is a subset of a dataset collected in early 2008 by Liu et al.~\cite{datasetA}.  Dataset B consists of 1,536 questions and is a subset of a dataset crawled in October 2008 by Aji and Agichtein~\cite{datasetB}.
We use subsets of the originally collected data because we only consider questions that were closed by the asker in less than 100 hours.
For each question in these datasets we know the time the question was posted, the arrival time of each answer to the question, and the time that the asker closed the question by selecting the best answer.

\begin{figure}
\begin{center}
		\includegraphics[width=0.5\textwidth]{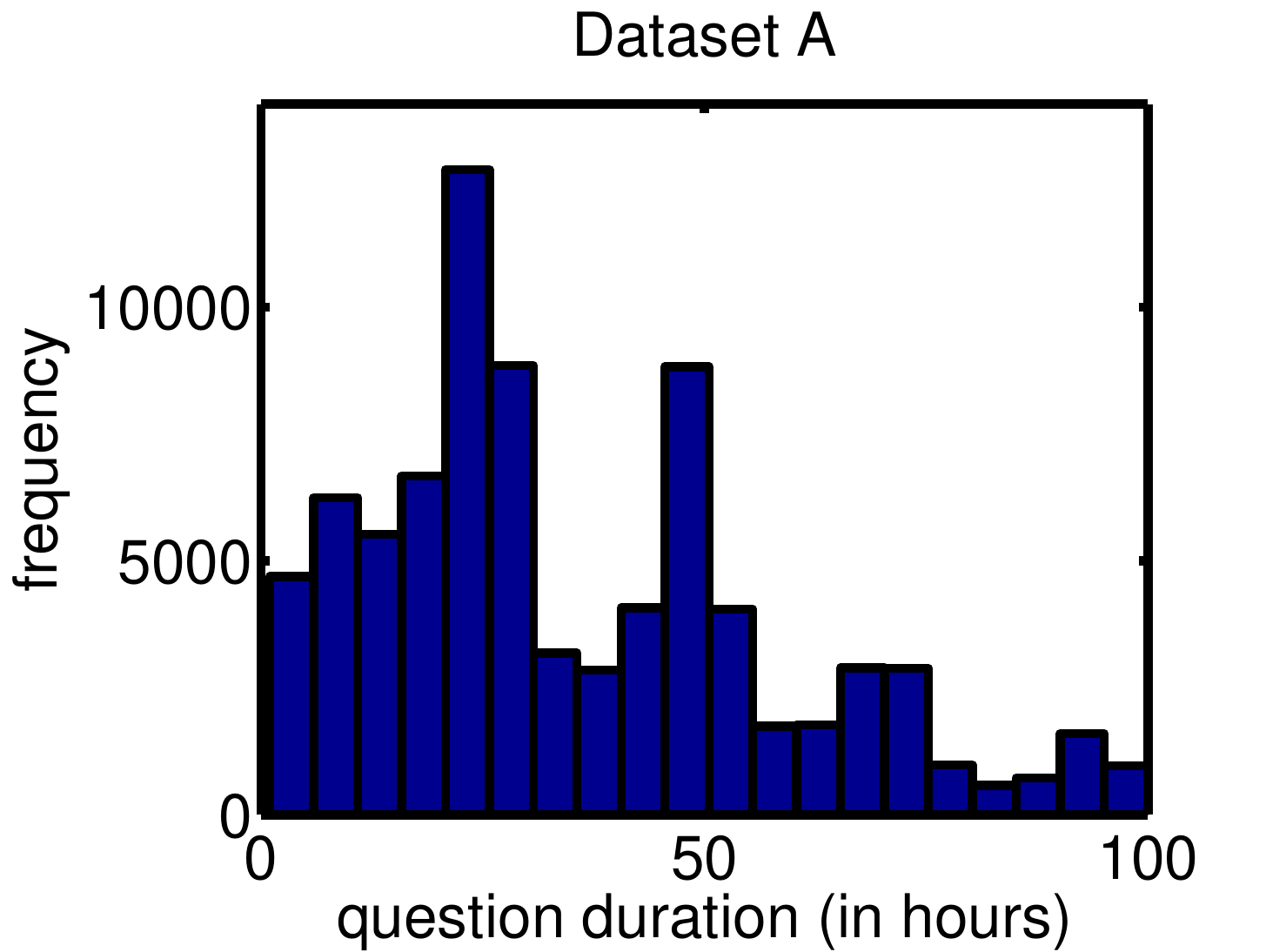}
\vspace{0.2cm}
		\includegraphics[width=0.5\textwidth]{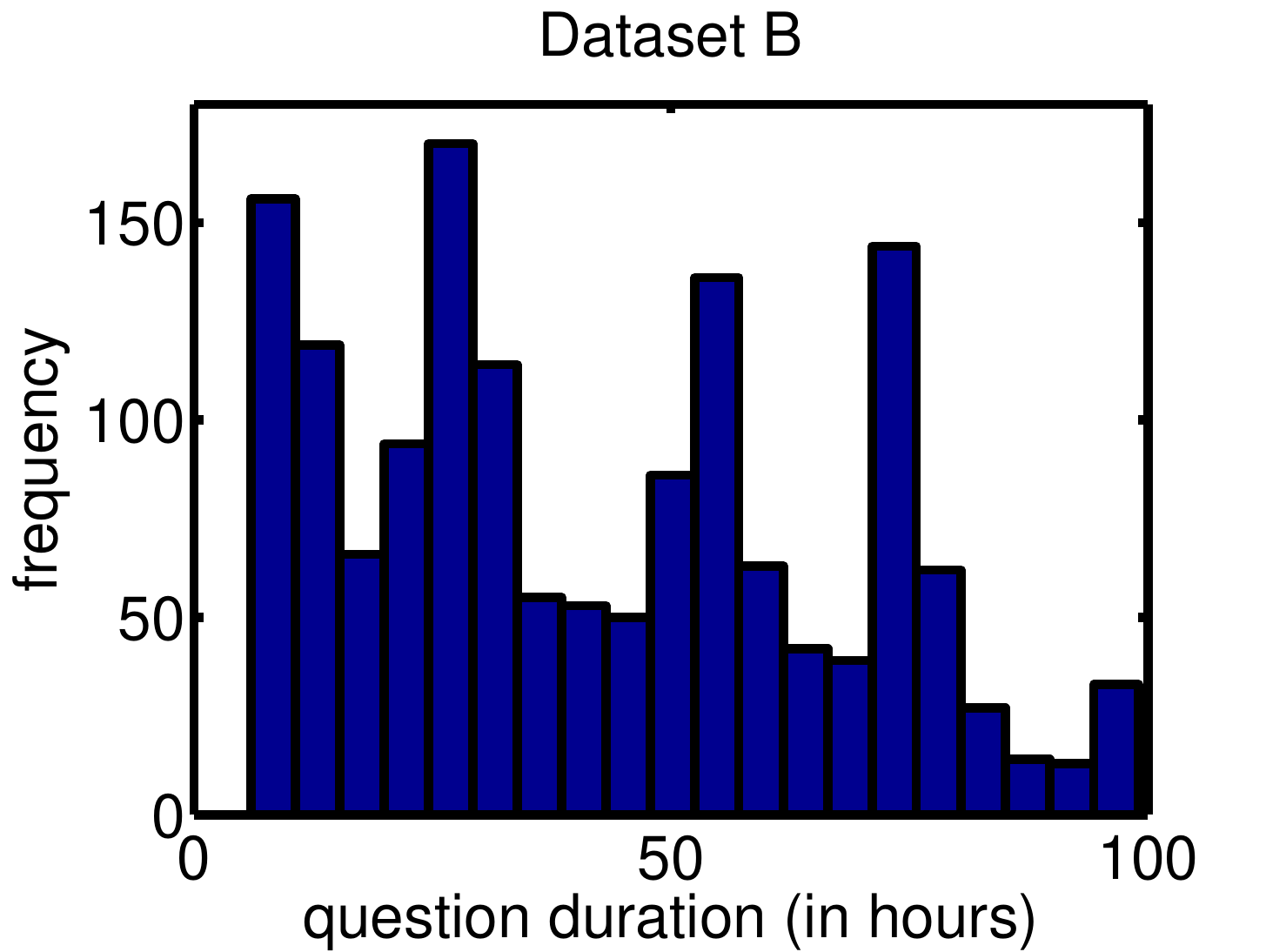}
\end{center}
\vspace{-1ex}
\caption
{Histograms of the number of hours that questions were open (before the asker chose the best answer) for Dataset A and B.}
\vspace{-2ex}
\label{fig:stop}
\end{figure}

One could argue that there is no reason for a user to close his question before the 4-day open period is over.  In particular, he could use the information from the answers he has received up to now, and still wait until the four days are over.  However, users often want to get a final answer and not have to rethink the question again.
Figure \ref{fig:stop} shows histograms of the number of hours that questions were open (before the asker chose the best answer).  Notice that a high percentage of questions closes within one day after the question was posted: 38\% for Dataset A and 29\% for Dataset B.

\section{Speed-Accuracy Tradeoff}
\label{sec:tradeoff}

There are various reasons for which users post questions at \YAc.  Users may need help in performing a task, seek advise or support, or may be merely trying to satisfy their curiosity about some subject.  In any case, a user is trying to use information from the answers he receives to build his own final answer to his question.  A user wants to get an accurate final answer without waiting too long.  The accuracy of the user's final answer is subjective and hard to measure.  In this section we use the number of answers as an approximation of accuracy.  Thus, we expect that a user's utility increases in the total number of answers that his question receives, and decreases in the time he waits for answers to arrive.




Section \ref{sec:utility} develops our hypotheses drawing on a utility model.  In Section \ref{sec:elapsed} we test our first hypothesis.  In Section \ref{sec:logit} we introduce a discrete choice model, which we estimate in Section \ref{sec:discrete} to test the remaining hypothesis.  Finally, in Section \ref{sec:ut_est} we discuss the form of the utility function.

\subsection{Utility Model}
\label{sec:utility}

Let $n$ be the total number of answers at the time the user builds his final answer.  We assume that the user gets utility $u(n)$. Furthermore, we assume that the user incurs a cost $c(t)$ for waiting for time $t$.  Thus, the user is seeking to maximize $u(n) - c(t)$.

Suppose that $n$ answers have arrived.  The user can either terminate his search by choosing the best answer, or wait for additional answers.  If he terminates his search now, he can build his final answer using the $n$ answers that he has received, and thus get utility $u(n)$.  If he chooses to wait and a new answer arrives $t$ time units later, then he will have $n+1$ answers, but will have also incurred a cost $c(t)$ for waiting.  His utility will then be $u(n+1) - c(t)$. The user is better off stopping if $u(n+1) - u(n) < c(t)$, and continuing if $u(n+1) - u(n) > c(t)$.
In words, {\em the user decides to close the question if the cost of waiting for one more answer exceeds the incremental benefit}.
On the other hand, {\em the user decides to wait for one more answer if the cost of waiting is smaller than the incremental benefit of having one more answer}.

Our previous description assumes that the user knows when the next answer will arrive, which is not the case in reality.  More realistically, we can assume that the user uses an estimate to calculate his cost.  Thus, the user closes the question if $u(n+1) - u(n) < c(\tau)$, and waits for the next answer if $u(n+1) - u(n) > c(\tau)$, where $\tau$ is now the user's estimate on how long he will have to wait until the next answer arrives.
More generally, let $T$ be a random variable that describes the user's belief on how long it will take until the next answer arrives.  Then, the user is better off closing the question if $u(n+1) - u(n) < E[c(T)]$, and continuing if $u(n+1) - u(n) > E[c(T)]$.

The strategy we just described is myopic, since it assumes that a user decides whether to wait (i.e., not to close the question) by only considering whether he is better off waiting for one more answer.  
Alternatively, if the user knew when each answer is going to arrive in the future, we could consider a global optimization problem: if the $i$-th answer is expected to arrive at time $t_i$, the user would choose to close the question at the time $t_j$ that maximizes $u(j) - c(t_j)$.
However, in the context of \YA it is impossible for users to know when all future answers will arrive.  It is thus more realistic to assume that users myopically optimize as randomness is realized.

\begin{figure}
\begin{center}
\begin{tabular}{|l|}
  \hline
If $u(n+1) - u(n) < E[c(T)]$, then close the question\\
\hline
If $u(n+1) - u(n) > E[c(T)]$, then wait \\
  \hline
\end{tabular}
\end{center}
\caption{Myopic decision rule.}
\label{fig:myopic}
\end{figure}

We summarize the myopic decision rule in Figure \ref{fig:myopic}.
It implies that a user is more likely to close the question when $u(n+1) - u(n)$ is small and/or $E[c(T)]$ is large.

We next develop our hypotheses building on the myopic decision rule.
Our hypotheses can be grouped in two categories.  The first category is based on the assumption that the marginal benefit of having one more answer decreases as more answers arrive; the second considers how users estimate when the next answer will arrive.

The user's valuation for having $n$ answers is $u(n)$.  We expect that $u(n)$ is concave, i.e., the marginal benefit of having one more answer decreases as the number of answers increases.  According to the myopic decision rule (Figure \ref{fig:myopic}), the user is more likely to close his question when $u(n+1) - u(n)$ is small.  Since we expect that $u(n+1) - u(n)$ is decreasing in $n$, the user is more likely to close his question when $n$ is large, i.e., when he has already received a large number of answers.   We test this in two ways, outlined in Hypotheses \ref{h:elapsed} and \ref{h:number}.

\begin{hypothesis}
\label{h:elapsed}
The amount of time that a user waits before closing his question is decreasing in the number of answers that the question has received.
\end{hypothesis}

\begin{hypothesis}
\label{h:number}
A user is more likely to close his question if the question has received many answers.
\end{hypothesis}

The user believes that the time until the next answer arrives is described by some random variable (which can be degenerate if he is only using an estimate).
It is reasonable to assume that a user forms his belief using the information available to him: the arrival times of previous answers and the time he has waited since the last answer arrived.

A particularly important summary statistic is the last inter-arrival time, i.e., the time between the arrivals of the two most recent answers.  The last inter-arrival time is an estimate of the inverse current arrival rate of answers.  Thus, the user may use the last inter-arrival time as an estimate of the next inter-arrival time, i.e., the time between the arrival of the last answer and the next answer.  More generally, the user may form a belief on the next inter-arrival time that depends on the last inter-arrival time on some increasing fashion.
Then, if the last inter-arrival time is large, the user expects to wait a long time until he receives another answer, thus incurring a large waiting cost.  This encourages the user to close the question now.  This is the context of Hypothesis \ref{h:interarrival}.

\begin{hypothesis}
\label{h:interarrival}
A user is more likely to close his question if the last inter-arrival time is large.
\end{hypothesis}

This hypothesis is based on the assumption that the last inter-arrival time may be used as an estimate for the next inter-arrival time.
However, if a long time has elapsed since the last answer arrived (e.g., a longer period than the last inter-arrival time), the user becomes less certain about this estimate.  The increased uncertainty may lead him to expect a greater waiting cost until the next answer arrives.  In turn, this encourages the user to close the question, as is outlined in Hypothesis \ref{h:hours}.

\begin{hypothesis}
\label{h:hours}
A user is more likely to close his question if a long time has elapsed since the most recent answer arrived.
\end{hypothesis}

Hypothesis \ref{h:elapsed} is tested in Section \ref{sec:elapsed}.  Then, in Section \ref{sec:logit} we introduce a discrete choice model, which we estimate in Section \ref{sec:discrete} to test Hypotheses \ref{h:number}, \ref{h:interarrival}, and \ref{h:hours}.

\subsection{Time Between Last Arrival and Closure}
\label{sec:elapsed}

In this section we test whether a user waits longer before closing his question when the question has received a small number of answers (Hypothesis \ref{h:elapsed}).

For every question we consider the following variables:
\begin{itemize}
\item TotalAnswers: the total number of answers that the question received.  This is the number of answers at the time that the asker closed the question.
\item ElapsedTime: the time between the arrival of the last answer and the time the user closed the question.
\end{itemize}
We test for correlation between TotalAnswers and ElapsedTime.
The results are presented in Table \ref{tab:cor}.
For both datasets, we find that TotalAnswers and ElapsedTime are negatively correlated, bringing support for Hypothesis \ref{h:elapsed}.

\begin{table}
\begin{center}
\begin{tabular}{|c|c|c|}
  \hline
   & Dataset A & Dataset B\\
   \hline
  Correlation coefficient   &  -0.126*** &  -0.148***\\
  95\% conf interval &  [-0.132, -0.119] &  [-0.196, -0.098]\\
  \hline
  Observations &  81,832 &  1,536\\
  \hline
\end{tabular}
\end{center}
\caption{Correlation between the number of answers (TotalAnswers) and the time that the user waits before closing the question (ElapsedTime).  *, ** and *** denote significance at 1\%, 0.5\% and 0.1\% respectively.}
\label{tab:cor}
\end{table}

\begin{figure}
\begin{center}
		\includegraphics[width=0.5\textwidth]{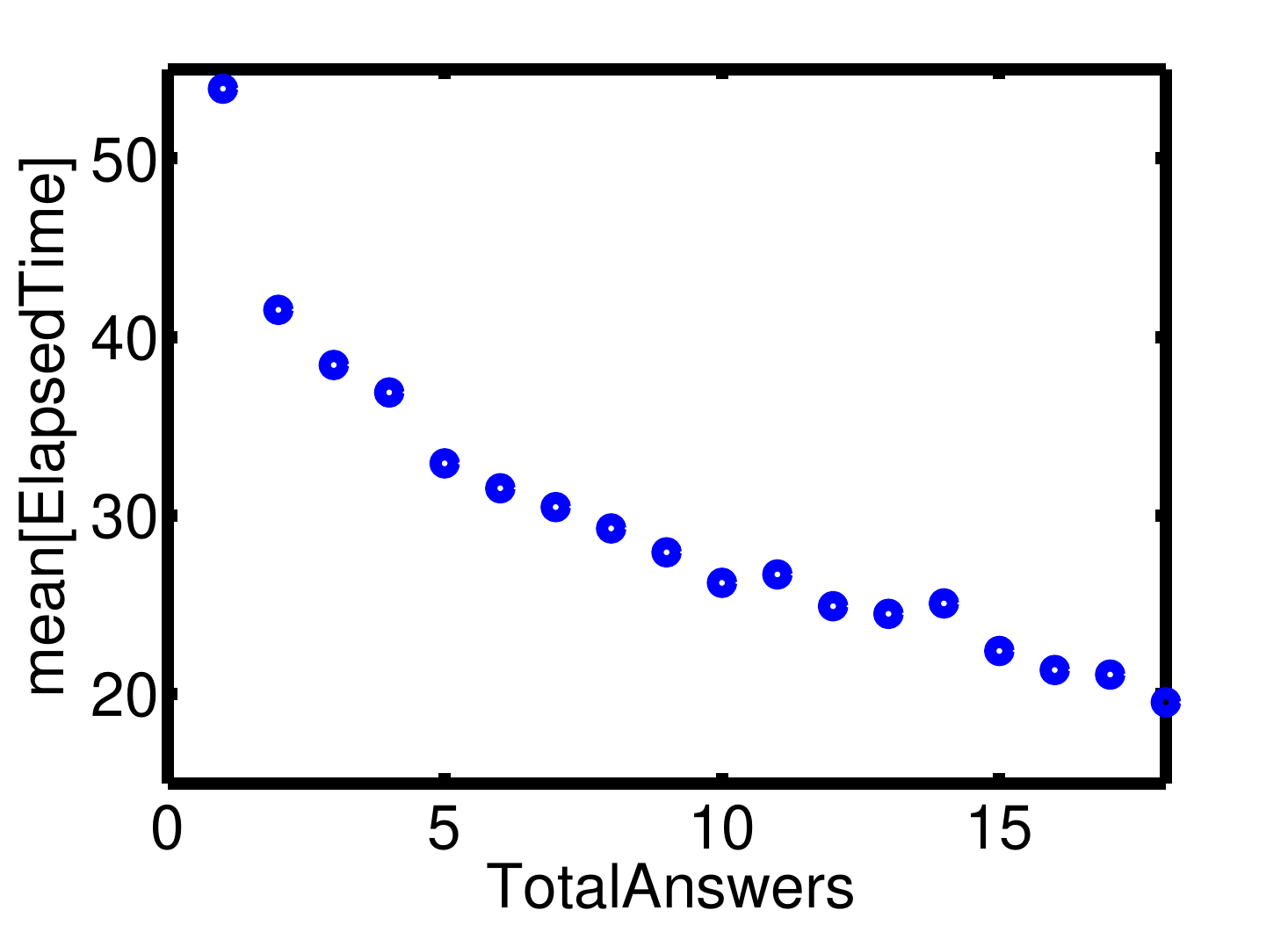}
\end{center}
\vspace{-1ex}
\caption
{Mean elapsed time between the time that the last answer arrived and the time that the question was closed (in hours) for Dataset A.
The horizontal axis shows the total number of answers at the time that the question was closed.}
\vspace{-2ex}
\label{fig:elapsed_time}
\end{figure}

Figure \ref{fig:elapsed_time} shows the mean elapsed time between the time that the last answer arrived and the time that the user chose the best answer in hours (i.e., the mean value of ElapsedTime) as a function of the total number of answers at the time (TotalAnswers).\footnote{We do not include questions with more than 18 answers in Figure \ref{fig:elapsed_time}, because for popular questions we have less observations in the dataset, resulting in more noise in the mean value of ElapsedTime.  We note however that the correlation test of Table \ref{tab:cor} is based on all the questions in the datasets.}
We observe that the time between the last answer and the closure of the question decreases as the number of answers increases.
For instance,  users wait more than 50 hours on average before closing their questions when they have only received one answer, while they wait less than 35 hours on average before closing their questions when they have received five answers.

Both Table \ref{tab:cor} and Figure \ref{fig:elapsed_time} suggest that the user is willing to wait more (and incur more cost from waiting) for an additional answer if only a few answers have arrived up to now.  This implies that the marginal benefit of having one additional answer decreases as the number of answers increases.

\subsection{Model Specification}
\label{sec:logit}

In this section we introduce a logit model, which is estimate in Section \ref{sec:discrete} to test Hypotheses \ref{h:number}, \ref{h:interarrival}, and \ref{h:hours}.

A user posts a question.  Then, at various points in time he revisits \YA to see the answers that his question has received, and decides whether to close the question by selecting the best answer.  We are interested in the probability that the user closes the question during a given visit.
For every visit we consider the following variables:
\begin{itemize}
\item $p$: the probability that the user closes the question during the visit.
\item $n$: the number of answers that the question has received by the time of the visit.
\item $l$: the last inter-arrival time, i.e., the time between the arrivals of the two most recent answers (at the time of the visit).
\item $w$: the time since the last answer arrived, i.e., the time that the user has been waiting for an answer since the last arrival.  This is equal to the difference between the time of the visit and the arrival time of the most recent answer.
\end{itemize}
The number of observations per question depends on the inter-arrival times between answers.

Recall the utility model introduced in Section \ref{sec:utility}.  If the user closes the question at $n$ answers, his utility is $u(n)$.  Suppose that the user believes that the next answer will arrive in time $T$, where $T$ is some random variable.  Moreover, we assume that the user uses the last inter-arrival time $l$ and the time since the last answer $w$ to form his belief; that is $T$ depends on $l$ and $w$, and we write $T(l,w)$.
Then, the user expects to obtain utility $u(n+1) - E[c(T(l,w))]$ from waiting.  According to the myopic decision rule (Figure \ref{fig:myopic}), the user decides whether to close the question or not depending on which of the expressions $u(n)$, $u(n+1) - E[c(T(l,w))]$ is larger.

We now perturb $u(n)$ and $u(n+1) - E[c(T(l,w))]$ with some noise.  In particular, we assume that the user's utility is
\[u(n) + \epsilon_0\]
if he closes the question, and
\[u(n+1) - E[c(T(l,w))] + \epsilon_1\]
if he waits for the next answer.  Suppose that $\epsilon_0$ and $\epsilon_1$ are assumed to be independent type 1 extreme value distributed.  Then, the difference $\epsilon_1 - \epsilon_0$ is logistically distributed.  Thus, the probability of closing the question is
\begin{align*}
p &= Pr[u(n) + \epsilon_0 > u(n+1) - E[c(T(l,w))] + \epsilon_1] \notag \\
         &= Pr[\epsilon_1 - \epsilon_0 < E[c(T(l,w))] - (u(n+1) - u(n))] \notag\\
         &= \Lambda(E[c(T(l,w))] - (u(n+1) - u(n))),
         \label{eq:logut}
\end{align*}
where
\[\Lambda(z) = \frac{1}{1+e^{-z}}\]
is the logistic function.

The previous argument gives rise to the logit model, a standard discrete choice model in microecomics (see e.g., \cite{econometrics}).

In Section \ref{sec:discrete} we estimate the following model:

\begin{equation}
\label{eq:logit}
p = \Lambda(\alpha + \beta_1 n + \beta_2 l + \beta_3 w),
\end{equation}
so that
\[E[c(T(l,w))] - (u(n+1) - u(n)) = \alpha + \beta_1 n + \beta_2 l + \beta_3 w.\]
This implies that the marginal benefit of having one more answer when $n$ answers have arrived is
\begin{equation}
\label{eq:utility}
u(n+1) - u(n) = \alpha_u - \beta_1 n
\end{equation}
and the expected cost of waiting for the next answer is
\begin{equation}
\label{eq:cost}
E[c(T(l,w))] = \alpha_c + \beta_2 l + \beta_3 w
\end{equation}
such that
\[\alpha_c - \alpha_u = \alpha.\]
Equations \eqref{eq:utility} and \eqref{eq:cost} are used in Section \ref{sec:ut_est} to interpret the estimated parameters of \eqref{eq:logit}.

\subsection{Model Estimation}
\label{sec:discrete}

In this section we use logistic regression to estimate \eqref{eq:logit}, i.e., we estimate the probability that a user closes his question as a function of (i) the number of answers ($n$), (ii) the last inter-arrival time ($l$), and (iii) the time that the user has waited since the last answer arrived ($w$).  We find that the probability of closing the question increases with all three variables, supporting Hypotheses \ref{h:number}, \ref{h:interarrival}, and \ref{h:hours} respectively.

We estimate \eqref{eq:logit} assuming that users visit \YA to check for new answers to their question every hour after the last answer arrived.
The maximum likelihood estimators are given in Tables \ref{tab:logit20} and \ref{tab:logitBall}.  All parameter estimates are statistically significant at the 0.001 level.  We also used a generalized additive model~\cite{learning} to fit the data, which suggested that the assumed linearity in \eqref{eq:logit} is a good fit for the data.

It is worth noting that our results do not heavily depend on our assumption that users check for new answers every hour.  We get similar estimates for $\beta_1$, $\beta_2$ and $\beta_3$, if we assume that users check for answers every 2 hours, every 5 hours, or every 30 minutes.  For instance, if we assume that users check for new answers every 2 hours, we get
\[(\beta_1, \beta_2, \beta_3) = (0.026, 0.027, 0.022)\]
instead of
\[(\beta_1, \beta_2, \beta_3) = (0.027, 0.028, 0.021)\]
for Dataset B.

Due to a sampling bias, Dataset A contains disproportionately many questions with more than 20 answers.  To decrease the effect of this bias in the logistic regression we only considered questions with less than 20 answers for Dataset A.  For Dataset B we run two logistic regressions: in one we only use questions with less than 20 answers (for the sake of comparison with Dataset A in Table \ref{tab:logit20}); in the other we consider the whole dataset.  The latter is shown in Table \ref{tab:logitBall}.

We observe that the estimates we get for the two datasets are close to each other.  
This suggests that our regressions are capturing how users behave with respect to when they close their questions, and that this behavior did not significantly change in the months between the times that the two datasets were collected.

\begin{table}
\begin{center}
\begin{tabular}{|c|c|c|}
  \hline
    & Dataset A &  Dataset B \\
  \hline
  $\alpha$  &-4.006*** (0.0116) &  -4.469*** (0.0064)\\
  $\beta_1$   & 0.031*** (0.0016)  &   0.036*** (0.0006)\\
  $\beta_2$ & 0.025***  (0.0005)&   0.029*** (0.0002)\\
  $\beta_3$     & 0.018*** (0.0002) &   0.021*** (0.0001)\\
  \hline
  Observations& 1,104,568 & 53,413 \\
  \hline
\end{tabular}
\end{center}
\caption{The effect of the number of answers, the last inter-arrival time, and the time since the last answer on the probability of closing the question with $p$ as the dependent variable.  The questions of Datasets A and B with less than 20 answers are considered in these regressions. *, ** and *** denote significance at 1\%, 0.5\% and 0.1\% respectively. Standard errors are given in parenthesis.}
\label{tab:logit20}
\end{table}

\begin{table}
\begin{center}
\begin{tabular}{|c|c|}
  \hline
   & Estimate\\
  \hline
  $\alpha$  &-4.408*** (0.0603)\\
  $\beta_1$ & 0.027*** (0.0005)\\
  $\beta_2$ & 0.028*** (0.0002)\\
  $\beta_3$ & 0.021*** (0.0001)\\
  \hline
  Observations & 54,914\\
  \hline
\end{tabular}
\end{center}
\caption{The effect of the number of answers, the last inter-arrival time, and the time since the last answer on the probability of closing the question with $p$ as the dependent variable for Dataset B.  *, ** and *** denote significance at 1\%, 0.5\% and 0.1\% respectively.   Standard errors are given in parenthesis.}
\label{tab:logitBall}
\end{table}

We can now draw qualitative conclusions by considering the signs of the estimated coefficients.
We use the fact that the sign of a coefficient gives the sign of the corresponding marginal effect (since the logistic function is increasing).

First, the probability of closing the question is greater when more answers have arrived, which supports Hypothesis \ref{h:number}.  This implies that the marginal benefit of having one additional answer decreases as the number of answers increases, and is consistent with Table \ref{tab:cor} and Figure \ref{fig:elapsed_time} of Section \ref{sec:elapsed}.

Second, the probability of closing the question is greater when the last inter-arrival time is greater, which supports Hypothesis \ref{h:interarrival}.  The inverse of the inter-arrival time gives the rate at which answers arrive.  Thus, when the last inter-arrival time is large, i.e., there is a time gap between the last answer and the answer before it, the user may expect that he will have to wait a long time until he receives the next answer.  This perceived high cost of waiting may encourage the user to close the question sooner when the last inter-arrival time is large.

Third, the probability of closing the question is greater when more time has elapsed since the last answer, which supports Hypothesis \ref{h:hours}.  As more time elapses since the last answer, the uncertainty increases, since the user does not know when the next answer will arrive.
The increased uncertainty may lead him to expect a greater waiting cost until the next answer arrives.  In turn, this encourages the user to close the question.

\subsection{Utility and Cost}
\label{sec:ut_est}

The following lemma establishes a specific quadratic form for the utility function.
\begin{lemma}
\label{l:utility}
If \eqref{eq:utility} holds, then
\begin{equation}
\label{eq:ut}
u(n) = \left(\alpha_u + \frac{\beta_1}{2}\right) n - \frac{\beta_1}{2} n^2 + u(0).
\end{equation}
\end{lemma}

\begin{proof}
If \eqref{eq:utility} holds, then
\begin{align*}
u(n) 
     &=u(0)+\sum_{i=0}^{n-1} (\alpha_u - \beta_1 i) \\
     &= \left(\alpha_u + \frac{\beta_1}{2}\right) n - \frac{\beta_1}{2} n^2 + u(0)
\end{align*}
\end{proof}

We observe that the utility function given in \eqref{eq:ut} is concave on $[0, \infty)$ for any value of $\alpha_u$ as long as $\beta_1 > 0$, which is the case for all the logistic regressions we run in Section \ref{sec:discrete}.
Moreover, for any fixed $\beta_1 > 0$ and $\alpha_u > 0$, the utility is unimodal: it is initially increasing (for $n < \lfloor\alpha_u/\beta_1 +0.5\rfloor$) and then decreasing (for $n > \lceil\alpha_u/\beta_1 +0.5\rceil$).  The latter may occur due to information overload; after a very large number of answers, the benefit of having one more answer may be so small that the cost of reading it exceeds the benefit, thus creating a disutility to the user.\footnote{The myopic decision rule is consistent with a utility function $u(n)$ that is decreasing for large values of $n$.  For those values of $n$ the cost of waiting clearly exceeds the benefit, and the user decides to close the question.}
Nevertheless, since questions at \YA rarely get a very large number of answers, the utility function given by \eqref{eq:ut} may be increasing throughout the domain of interest if $\alpha/\beta_1$ is sufficiently large.

\begin{figure}
\begin{center}
		\includegraphics[width=0.5\textwidth]{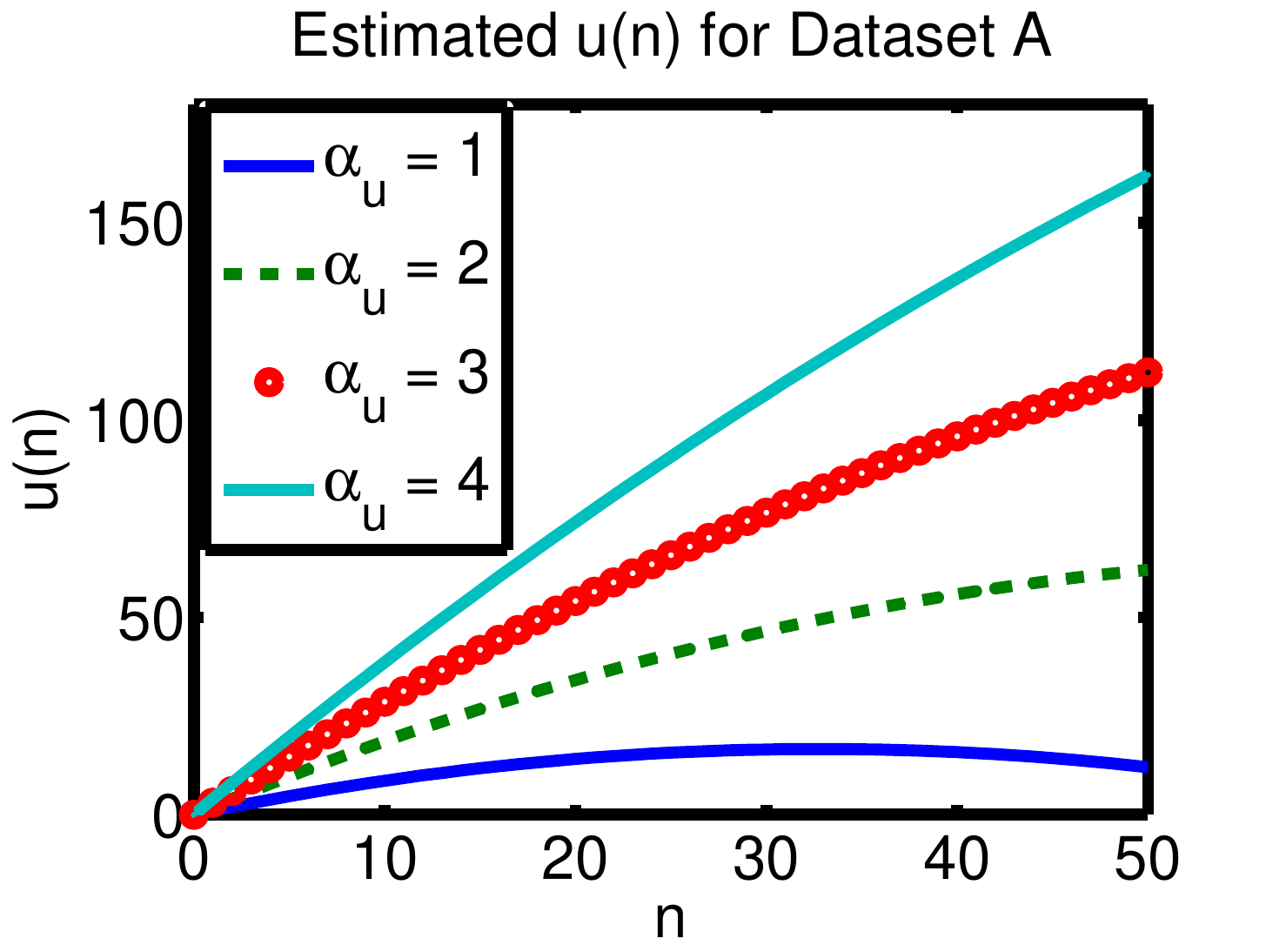}
\end{center}
\vspace{-1ex}
\caption
{Estimated $u(n)$ from Dataset A for $\alpha_u \in \{1,2,3,4\}$.}
\vspace{-2ex}
\label{fig:utility}
\end{figure}

We note that from \eqref{eq:logit} we estimate $\beta_1$ and $\alpha$, but cannot estimate $\alpha_u$.
For the sake of illustration we plot the estimated utility $u(n)$ from Dataset A (i.e., with $\beta_1 = 0.031$) for various values of $\alpha_u$ in Figure \ref{fig:utility}.  Since $\alpha_u = -\alpha + \alpha_c$ and in this case $\alpha \approx -4$, we consider $\alpha_u \in \{1,2,3,4\}$ in the plots; we also assume that $u(0) = 0$. A reasonable domain to consider is $[0,50]$, since questions rarely get more than 50 answers.
We observe that when $\alpha_u$ is small (e.g., $\alpha_u = 1$), then the estimated $u(n)$ is decreasing for large values of $n$ within the $[0,50]$ region; suggesting an information overload effect.  On the other hand, for $\alpha_u \in \{2,3,4\}$, the estimated utility function is increasing throughout $[0,50]$.  Moreover, as $\alpha_u$ increases, the curvature of the estimated utility decreases, something we can also conclude from \eqref{eq:ut}.

We next consider the cost of waiting.  Equation \eqref{eq:cost} suggests that the expected cost increases linearly in both the last inter-arrival time $l$ and the time since the last answer arrival $w$.  However, it is not possible to get a specific form for the function $c(t)$ in the way we did for $u(n)$ in Lemma \ref{l:utility}, because we do not have any information on $T(l,w)$.  If we make assumptions on $T(l,w)$, we can draw conclusions about $c(t)$.  For instance, if we assume that the user is using a single estimate $\tau(l,w)$ on the time until the next answer arrives, then \eqref{eq:cost} implies that
\[c(\tau(l,w)) = \alpha_c + \beta_2 l + \beta_3 w.\]
Moreover, if the estimate $\tau(l,w)$ is linear in $l$ and $w$ we conclude that the cost of waiting is linear.  On the other hand, a concave cost would be consistent with a convex estimate, and a convex cost would be consistent with a concave estimate.

\section{Assessing Quality}
\label{sec:quality}

Our previous analysis considers how the decision problem of the user depends on the number of answers and time.  There is a third aspect that affects the user's decision to close his question: the quality of the answers that have arrived up to now.  
In this section, we use an alternative model that incorporates quality, but does not incorporate time and the number of answers in the detail of Section \ref{sec:tradeoff}.  Our approach here is inspired by~\cite{surfing, surfing_science}.

Let $X_n$ be the value of the $n$-th answer.  This is in general subjective, and depends on the asker's interpretation and assessment. We assume that the value of an answer depends on both its quality and on the time that the user had to wait to get it.  For instance, $X_n$ may be negative if the waiting time was very large and the answer was not good (according to the user's judgement).
We model the values of the answers as a random walk, and assume that
\begin{equation}
\label{eq:RW}
X_{n+1} = X_n + Z_n,
\end{equation}
where the random variables $Z_n$ are independent and identically distributed.  For instance, if the user just got a high quality answer, he believes that the next answer will most likely also have high quality.  Similarly, if the user did not have to wait long for an answer, he expects that the next answer will probably arrive soon.  We note that \eqref{eq:RW} is consistent with the availability heuristic~\cite{availability}.

Every time that a user sees an answer, he derives utility that is equal to the answer's value.  We assume that the user discounts the value he receives from future answers according to a discount factor $\delta$.
Let $V(x)$ be the maximum infinite\footnote{It is straightforward to extend the results to a finite horizon problem.} horizon value when the value of the last answer is $x$.  Then,
\[V(x) = x  + \max\{0, \delta \cdot E(V(x+Z))\}.\]
In particular, the user decides to close the question if the value of closing exceeds the value of waiting for an additional answer.  If he closes the question, the user gets no future answers, and thus gets future value equal to 0.  On the other hand, if the user does not close the question, he gets value $E(V(x+Z))$ in the future, which he discounts by $\delta$.  Depending on which term is greater, the user decides whether to close the question or not.

We observe that $V(x)$ is increasing in $x$, which implies that $E(V(x+Z))$ is increasing in $x$.  We conclude that there exists a threshold $x^*$ such that it is optimal for the user to stop (i.e., close the question) when the value of the last answer is smaller than $x^*$ and to continue when the value of the last answer is greater than $x^*$.  The threshold $x^*$ satisfies $E(V(x^*+Z)) = 0$.

From an initial answer value, the user waits for additional answers, with values following a random walk as specified by \eqref{eq:RW}, until the value of an answer first hits the threshold value.  Thus, the number of answers until the user terminates the search is a random variable.  In the limit of true Brownian motion, the first passage times are distributed according to the inverse Gaussian distribution. Then, the probability density of the number of answers to a question is given by
\begin{equation}
\label{eq:invG}
f(x) = \sqrt{\frac{\lambda}{2 \pi}} x^{-3/2} \exp\left(-\frac{\lambda}{2 \mu^2 x}(x-\mu)^2\right),
\end{equation}
where $\mu$ is the mean and $\lambda$ is a scale parameter.  We note that the variance is equal to $\mu^3/\lambda$.

\begin{figure}
\begin{center}
		\includegraphics[width=0.5\textwidth]{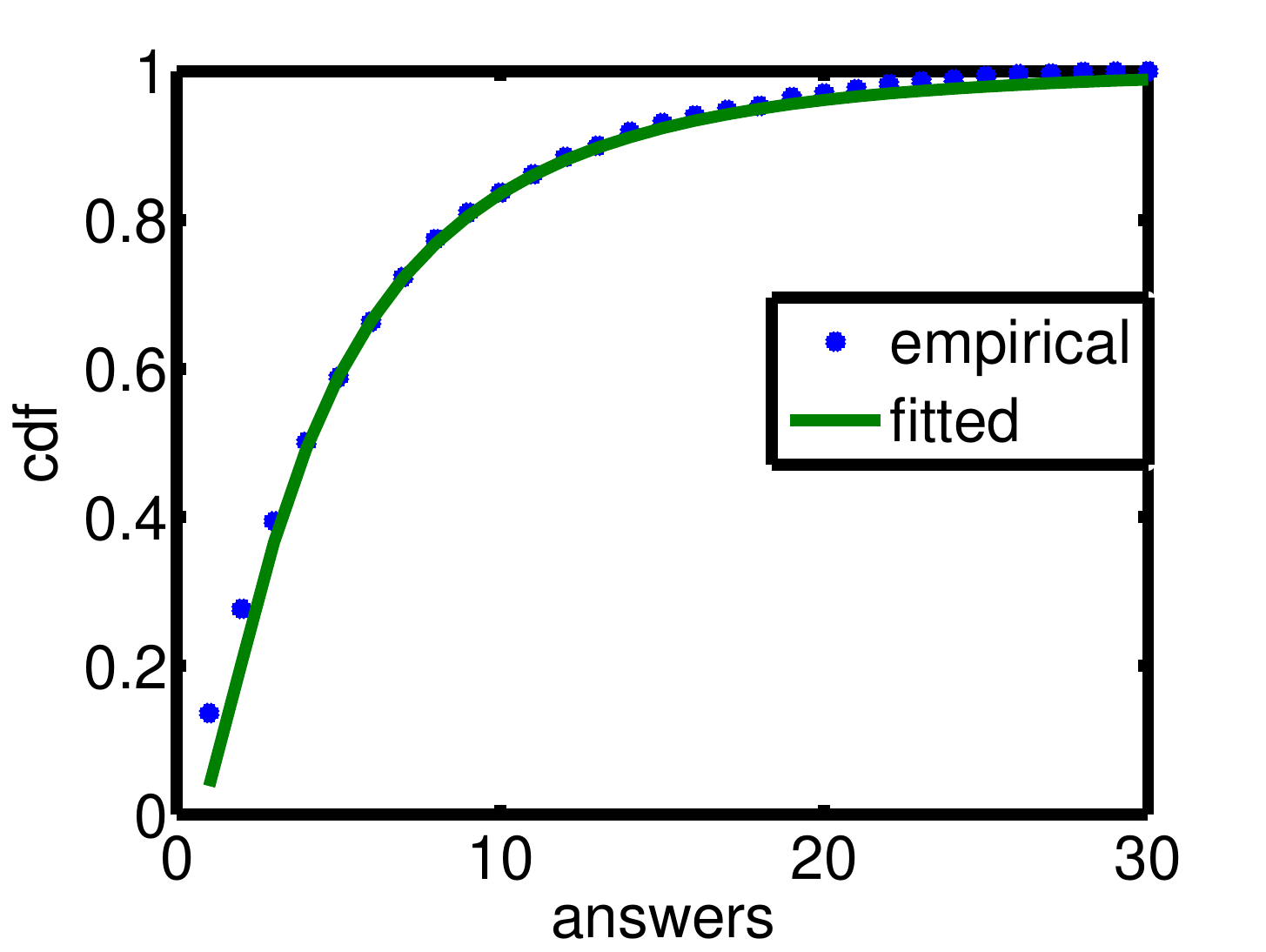}
\end{center}
\vspace{-1ex}
\caption
{Empirical and inverse Gaussian fitted cumulative distributions for Dataset B.
The points are the empirical cumulative distribution function of the number of answers.  The curve is the cumulative distribution function of the maximum likelihood inverse Gaussian.}
\vspace{-2ex}
\label{fig:cdf}
\end{figure}

We use Dataset B to test the validity of \eqref{eq:invG}.  We find that the maximum likelihood inverse Gaussian has $\mu = 6.1$ and $\lambda = 5.8$.  Figure \ref{fig:cdf} shows the empirical and fitted cumulative distribution functions.  We observe that the inverse Gaussian distribution is a very good fit for the data.

\begin{figure}
\begin{center}
        \includegraphics[width=0.5\textwidth]{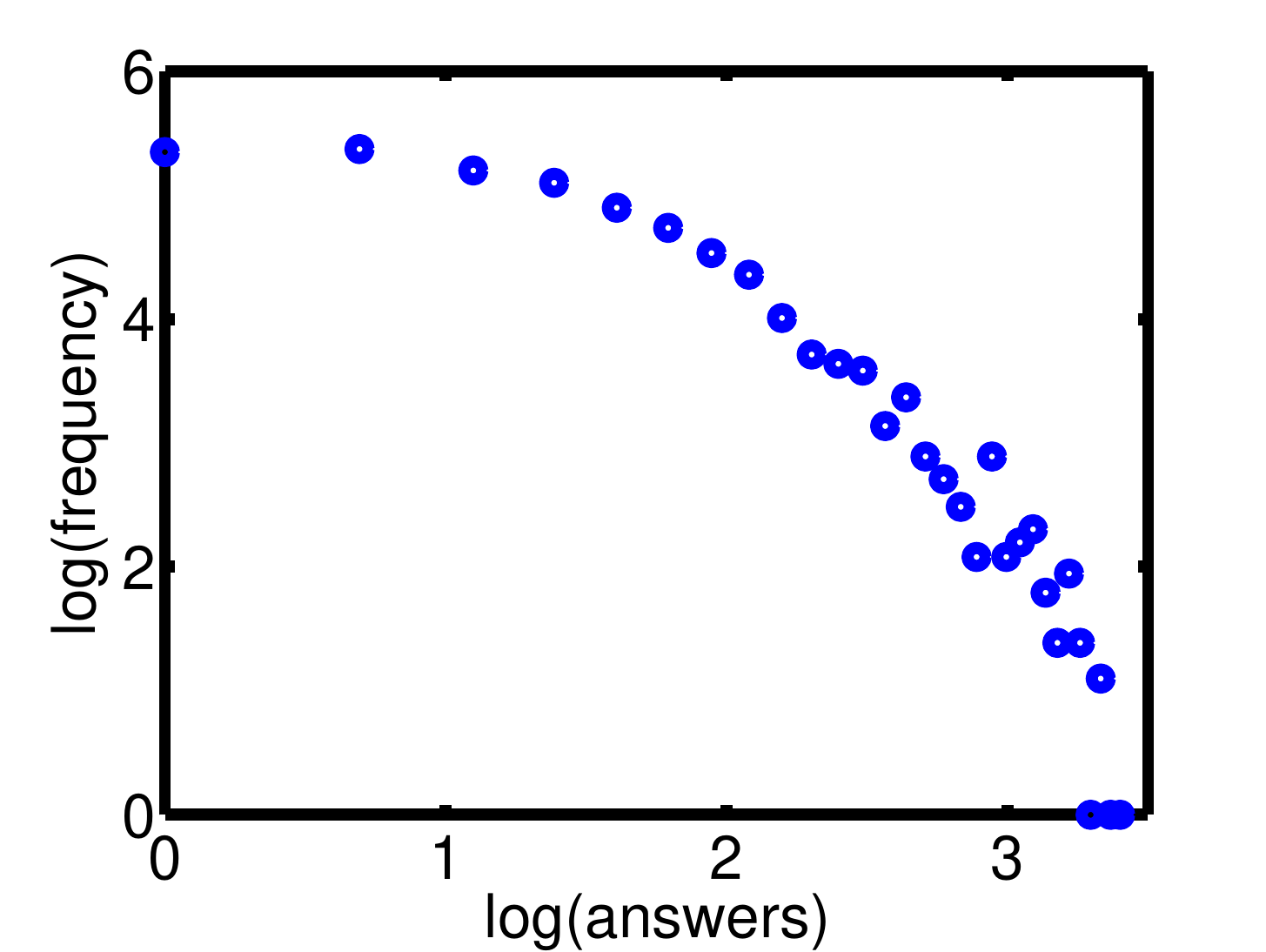}
\end{center}
\vspace{-1ex}
\caption
{The frequency distribution of the number of answers on log-log scales for Dataset B.}
\vspace{-2ex}
\label{fig:loglog}
\end{figure}

An important property of the inverse Gaussian distribution is that for large variance, the probability density is well-approximated by a straight line with slope -3/2 for larger values of $x$ on a log-log plot; thus generating a Zipf-like distribution.  This can be easily seen by taking logarithms on both sides of \eqref{eq:invG}.  In Figure \ref{fig:loglog} we plot the frequency distribution of the number of answers on log-log scales.  We observe that the slope at the tail is approximately -3/2.

\section{Conclusion}
\label{sec:conclusion}

This paper empirically studies how people behave when they face speed-accuracy tradeoffs.
We have taken two complementary approaches.

Our first approach is to study the speed-accuracy tradeoff by using the number of answers as a proxy for accuracy.  In particular, we assume that the user approximates the accuracy of his final answer by the number of answers that his question gets.  Thus, the user faces the following tradeoff: he prefers more to less answers, but does not want to wait too long.  We analyze \YA data to identify and quantify this tradeoff.  We find that users are willing to wait longer to obtain one additional answer when they have only received a small number of answers; this implies  decreasing marginal returns in the number of answers, or equivalently, a concave utility function.  We then estimate the utility function from the data.

Our second approach focuses on how users assess the qualities of the individual answers without explicitly considering the cost of waiting.  We assume that users make a sequence of decisions to wait for another answer, deciding to wait as long as the current answer exceeds some threshold in value.  Under this model, the probability distribution for the number of answers that a question gets is an inverse Gaussian, which is a Zipf-like distribution.  We use the data to validate this conclusion.

It remains an open question how to combine these two approaches in order to study the speed-accuracy tradeoff by jointly considering the number of answers, their qualities, and their arrival times.

We conclude by noting that our results could be used by \YA or other question answering sites to prioritize the way questions are shown to potential answerers in order to maximize social surplus.
The key observation is that a question receives answers at a higher rate when is it shown on the first page at \YAc.  On the other hand, the rate at which answers are received also depends on the quality of the question.
Using appropriate information about these rates as well as the utility function estimated in this paper, the site can position open questions with the objective of maximizing the sum of users' utilities.

\section{Acknowledgements}

We gratefully acknowledge Eugene Agichtein for providing the datasets~\cite{datasetA, datasetB}, as well as detailed information on how the data was collected.

\bibliographystyle{abbrv}
\bibliography{outline}

\end{document}